# The Potential Importance of Non-Local, Deep Transport on the Energetics, Momentum, Chemistry, and Aerosol Distributions in the Atmospheres of Earth, Mars and Titan

Scot C. R. Rafkin
rafkin@boulder.swri.edu
Dept. of Space Studies
Southwest Research Institute
1050 Walnut Street, Suite 300
Boulder, CO 80503
720-240-0116

Submitted to Planetary and Space Science

October 15, 2010

Pages: 25

Figures: 0

Tables: 0

**Abstract**

A review of non-local, deep transport mechanisms in the atmosphere of Earth provides a good foundation for examining whether similar mechanisms are operating in the atmospheres of Mars and Titan. On Earth, deep convective clouds in the tropics constitute the upward branch of the Hadley Cell and provide a conduit through which energy, moisture, momentum, aerosols and chemical species are moved from the boundary layer to the upper troposphere and lower stratosphere. This transport produces mid-tropospheric minima in quantities such as water vapor and moist static energy and maxima where the clouds detrain. Analogs to this terrestrial transport are found in the strong and deep thermal circulations associated with topography on Mars and with Mars dust storms. Observations of elevated dust layers on Mars further support the notion that non-local deep transport is an important mechanism in the atmosphere of Mars. On Titan, the presence of deep convective clouds almost assures that non-local, deep transport is occurring and these clouds may play a role in global cycling of energy, momentum, and methane. Based on the potential importance of non-local deep transport in Earth's atmosphere and supported by evidence for such transport in the atmospheres of Mars and Titan, greater attention to this mechanism in extraterrestrial atmospheres is warranted.

## 1 Introduction

Non-local, deep transport (NLDT) plays a potentially important role in the dynamics, energetics, and chemistry of planetary atmospheres such as Mars and Titan, much as it does for Earth. NLDT is accomplished by transport within coherent and vertically extensive circulations and provides for both up- and down-gradient transport that, unlike diffusion, does not fundamentally depend on the local gradient. One of the most important aspects of NLDT is its ability to produce local maxima of various quantities in the absence of a local source, and it is this characteristic that makes it "non-local".

While NLDT is recognized as being important within the atmosphere of Earth, it has received little recognition in the atmospheres of other planets. This neglect is striking given observations of NLDT processes such as deep convection on Titan and Mars dust storms. The neglect is equally striking given observations of the tell-tale signs of NLDT in other atmospheres, for example the elevated dust layers on Mars.

If there is anything to be learned from NLDT on Earth and then applied to other planetary atmospheres, it is that it is fundamentally important to incorporate this process in order to develop a consistent and proper understanding of global and often regional energy, mass, momentum, chemical and aerosol cycles. Or, stated differently, any model or description of global atmospheric cycles that do not include NLDT will be in error in proportion to the importance of NLDT processes.

## 2 NLDT on Earth

The best known example of NLDT is that of "hot towers" first described by Riehl and Malkus (1958) to explain the upper tropospheric maximum of moisture and enthalpy (a.k.a. moist static energy) in the Earth's tropical atmosphere and a corresponding moist static energy minimum in the mid-troposphere (Ooyama 1969). The source of moisture and enthalpy is the warm tropical ocean and the adjacent atmospheric boundary layer. Diffusion and large-scale vertical motion cannot explain the local tropopause maximum. It is the rapid and deep transport through updraft cores of tropical thunderstorms—which Riehl and Malkus termed hot towers—that provide the mechanism by which the upper tropospheric maximum is produced and maintained. Importantly, mixing between the cloud cores and the cloud-free environment is limited so that moisture and enthalpy are nearly conserved during the transport.

Since the landmark paper by Riehl and Malkus over half a century ago, the recognition of the importance of NLDT through deep convective clouds has expanded beyond simply moisture and enthalpy. The transport through these clouds is now recognized as an integral part of the Hadley Cell with critically important effects on the global moisture and energy cycles. Momentum is transported through the clouds and this provides a mechanism by which planetary surface angular momentum can be injected at higher altitudes while depositing little angular momentum in between. The clouds are also responsible for transport of chemical species, such as $O_3$, and for the injection of aerosols into the upper troposphere and lower stratosphere.

### 2.1 The Hadley Cell and Global Moisture, Energy and Momentum Budgets

The large-scale radiative imbalance of Earth must be countered in order to maintain a quasi-equilibrium thermal state. Atmosphere and ocean circulations are the mechanism by which energy is transferred to counter the radiative imbalance. In the tropical atmosphere, energy is moved vertically before it is transported in the poleward branch of the Hadley Cell (Webster 2005). Within the rising branch of the Hadley Cell, latent heat of evaporation originating from the warm oceans is realized as sensible heat within thunderstorms. These storms occupy only a small fraction of the total area of the tropics. In contrast, the non-cloud environment occupies most of the tropics and it is within this broad area where compensating subsidence from convection is found. The compensating subsidence produces adiabatic warming, which is primarily how the sensible heat within the small fractional area of clouds is communicated to the rest of the tropical atmosphere (Yanai et al. 1973).

It is the net difference between the narrow but strong updrafts of convection and the weaker convective downdrafts and compensating subsidence that results in the rising branch of the Hadley Cell. The Hadley Cell, therefore, is a mathematical construct and not a physical circulation. Actual measurements of vertical mass flux in the tropics would reveal either strong upward motion in the clouds or more gentle but widespread downward motion outside of the clouds, but would not reveal anything resembling the mean rising branch of the Hadley Cell. Only once the net motion is considered does the Hadley Cell appear.

It follows then, that vertical advection is performed by actual physical circulations: either within clouds or in the surrounding subsidence. Vertical transport is not performed by the Hadley Cell, since it is not a physical circulation. This has important consequences. Advection by a slow and uniform Hadley cell would produce either a gradient of energy and moisture with a maximum at the surface, or it would produce a well-mixed troposphere of uniform energy and moisture. This is not observed.

What is observed is an atmospheric state that is consistent with transport by physical transport mechanisms. Namely, there is a maximum of moisture near the surface (where the source of water is) and at the top of the tropopause where clouds detrain (Jordan and Jordan 1954; Dunion and Marron 2008). The middle troposphere is remarkably dry. A similar structure is found for moist static energy: a maximum near the surface and in the upper troposphere and a mid-tropospheric minimum. It is, in fact, the mid-tropospheric minimum that makes that atmosphere unstable against moist convection.

The importance of considering the actual circulations as opposed to the mathematically constructed Hadley Cell goes beyond atmospheric structure. Some of the first experiments with general circulation models of the atmosphere investigated the effect of moist convection on dynamics. Manabe et al. (1970) switched off moist processes and compared the results to a control simulation with moist processes active. While the bulk circulation features were present in all simulations, the strength of the Hadley Cell was greatly reduced in the dry simulations and the trade winds were dramatically reduced. Eddy kinetic energy in the moist simulation compares well with observations in producing a maximum below the tropical tropopause and secondary maxima in the middle latitudes (associated with baroclinic storm systems). Not surprisingly, the tropical maximum in the dry simulation is almost completely absent. Deep convective transport in the tropics has global implications on the mean and eddy kinetic energy.

Momentum is also transported in convection. Wu et al. (2007) found substantial global impacts when momentum transport was included within convective parameterizations in general circulation modeling studies. In particular, equatorward flow at high latitudes was reduced and the intertropoical convergence zone was weaker. In a similar study that included convective momentum transport, Zhang and McFarlane (1995) found a strengthening of the Hadley Cell. Substantial global perturbations due to convective momentum transport were also reported by Richter and Rasch (2008). Therefore, NLDT of momentum by clouds provides an important source term in the global angular momentum budget.

Viewing the terrestrial general circulation holistically to include cloud circulations rather than just a Hadley cell alone leads to very different interpretations of water and energy transport. For example, neglecting the rapid and deep transport of mass, momentum, and energy by tropical thunderstorms and relying only on the Hadley cell for such transport leads to a grossly inaccurate representation of the hydrologic cycle (Donner al. 1982; Sud et al. 1992) and the Hadley cell itself. Similar effects were found by (Ose et al. 1989). Geller et al. (1988) showed that deep convection strongly influences the El Niño Southern Oscillation (ENSO), which is a dominant climate signal. In a comprehensive study of the effects of deep cumulus convection, Sud et al. (1992) found statistically significant effects on the circulation and on boundary layer and upper-level moisture structure, zonal winds, and moisture and heat flux transports.

**2.2 Chemistry**

Pommereau (2010) recently compiled balloon and aircraft observations from a variety of field experiments to determine that deep cloud transport has a major impact on upper tropospheric and stratospheric chemistry. This result is consistent with observations obtained by the Cloud-Aerosol Lidar and Infrared Pathfinder Satellite Observation (CALIPSO) mission (Vernier et al. 2009) all of which show that deep convective clouds have a substantial and important impact on the chemical composition of the upper troposphere and lower stratosphere. Ozone has been found to be a particularly good tracer of deep convective transport within clouds (e.g., Ching et al. 1988).

There is also a chemical-dynamical interaction within clouds such that the ventilation of boundary layer pollutants modifies chemistry within the clouds and has a strong influence on the concentration of species that are vented at higher altitudes. For example, Walcek et al. (1990) found almost a 50% increase on $O_3$ production rates due to the rapid transport of $NO_x$ within clouds. Liu et al. (1997) found not only large increases in $NO_x$ concentration due to clouds, but numerous other chemical species with reservoirs in the planetary boundary layer: e.g., $H_2O_2$, $O_3$, and $SO_2$. Numerous other studies (e.g., Wang and Prinn 2000, Kley et al. 1996, Kley 1997, and Alheit and Hauf 1994) all point to the importance of cloud transport in determining the distribution and concentration of chemical species.

It is important to once again emphasize that the transport accomplished by clouds cannot be accomplished or modeled through pure diffusive processes. Chemical models that attempt to reproduce observations using only diffusive transport will not be able to accurately do so in the cases where NLDT is important.

## 2.3 Aerosols

Aerosols, including those generated through forest fires, have been observed to undergo significant transport through deep convective clouds. The transport may be direct or it may involve condensation upon the aerosol, transport, and then evaporation that releases the aerosol at higher levels. Andreae et al. (2001) observed the transport of smoke from Brazilian rainforest fires into the upper troposphere and out over the Pacific Ocean. Numerous chemical species (CO, $CO_2$, acetonitrile, methyl chloride, hydrocarbons, NO, $O_3$) were also found within the smoke layer.

Flying at 7.5 km, aircraft in the INDOEX field campaign observed an aerosol concentration increase of up to 600 $cm^{-3}$, which rivals the background marine aerosol load (Engström et al. 2008), leaving little doubt that NLDT is an important mechanism in the global aerosol budget.

The earlier mentioned CALIPSO mission provides further support for the important role of NLDT of aerosols within clouds (Vernier et al. 2009).

### 2.4 Non-Cloudy NLDT

NLDT on Earth extends beyond that of tropical thunderstorms. NLDT occurs whenever tracers are transported through multiple atmospheric layers with little or no entrainment. The upward branch of the Asian Monsoon forced by the elevated topography of the Tibetan Plateau is a regional NLDT circulation with global implications. On a regional scale, sea-breezes and their interaction with topography produce a NLDT mechanism by which elevated layers of pollutants are produced in the Los Angeles basin, and probably other locations as well (Dacre et al. 2007). These elevated smog layers exhibit distinct layering each day, persist for many days, and are horizontally extensive. The photochemical smog cycle in L.A. is not complete and cannot explain observed pollutant concentrations without incorporation of NLDT. As with the NLDT in clouds, diffusive processes are insufficient to explain the reality of the situation.

Smoke and aerosols can be injected directly into the upper troposphere from the buoyant plumes of fires. This mechanism was found to be the dominant cause of thick elevated dust layers during the dry season in Africa (Schmid et al. 2003). Perhaps the best known large-scale aerosol phenomena is that of the elevated Asian dust plumes that periodically traverse the pacific. These dust layers originate from the rapid lofting of dust over Asia, usually by strong storm systems (e.g., Husar et al. 2001). Finally, as demonstrated by the complete halt of transcontinental traffic between North America and Europe following the eruption of Eyjafjallajökull in Iceland in 2010, volcanic plumes are a dramatic example of the global impact of a regional but persistent NLDT event.

## 3 NLDT in other Planetary Atmospheres

There is no reason to suspect that Earth is unique among planetary atmospheres with respect to NLDT. Therefore, it is reasonable to investigate to what extent NLDT may be present in other planetary atmospheres and to assess how important that process may be in global dynamics and chemistry. Discussion is restricted to the terrestrial planetary bodies of Mars and Titan, although NLDT may be important in the atmosphere of Venus, Jupiter, Saturn, Uranus and Neptune. With respect to Mars and Titan, recognition of NLDT in these atmospheres has been limited, if not non-existent, within the literature. However, as will be shown, there is sufficient observational evidence to suspect that NLDT does play an important role in the atmospheres of each of these bodies.

### 3.1 Mars

Elevated and detached haze layers were recognized in the Mariner television images (Anderson and Leovy 1978), Viking orbiter limb images (Jaquin et al. 1986), and images obtained by the Mars Orbiter Camera on the Mars Global Surveyor (MGS). Thermal Emission Spectrometer limb dust retrievals, also show elevated dust layers or "layers of altitude increasing dust" (Clancy et al. 2010). It should not be surprising that similar and persistent elevated dust layers have now been observed by the Mars Climate Sounder on the Mars Reconnaissance Orbiter (McCleese et al. 2010).

Despite the elevated layers of dust and water on Mars, the transport of these quantities has, with few exceptions, been credited to the Hadley Cell (e.g., Wilson et al. (2002), Kahre et al. (2006), and Basu et al. (2004)). The most widely used dust prescription for models—the Conrath-ν vertical distribution—rests on the notion of a balance between upward diffusion of dust and gravitational sedimentation (Conrath 1975). As noted for Earth, the mean meridional circulation (or diffusion) cannot produce elevated layers of water and dust, leading (McCleese et al. 2010) to state "the existence of this maximum suggests that current understanding of the mechanisms by which dust enters and leaves the atmosphere is incomplete." But, this is not entirely accurate.

While the details of mechanisms by which dust enters the atmosphere is incomplete, Rafkin et al. (2002) specifically identified NLDT as a likely mechanism by which elevated dust layers could be created, including the following statements:

> *"The flow becomes horizontally divergent near the top of the circulation, where the outflow branches of the circulation transport dust laterally up to a thousand or more kilometres on either side of the volcano…A dust particle of less than 1 um in diameter falling at 1 cm/s would require in excess of 20 Martian days to fall 20 km in altitude in the absence of vertical motion in the atmosphere. Such a particle could travel significant horizontal distances within this timeframe. Dust less than 1 um in diameter would remain aloft for even longer periods, and could easily circle the planet given only the moderate wind speeds predicted by Mars GCMs."*

The paper concludes with the following statement,

> "*Mars is dotted with numerous topographic features that are too small in horizontal extent to be captured by GCMs, but which may produce large-scale thermal circulations that could perturb the general circulation. Our results suggest that mesoscale thermal circulations may collectively be important in the atmospheric dust* *budget, and individually can produce strong regional perturbations in the background large-scale flow."*

Included in Rafkin et al. (2002) is a figure showing an elevated dust layer at 15-25 km, perhaps not coincidentally where one has now been observed by MCS.

The importance of NLDT for Mars was further emphasized in Rafkin (2003) with specific reference to the hot towers of Riehl and Malkus (1958) and the unambiguous claim that "The existence of deep thermal circulations forced by topography is almost certain on Mars." This claim was not referring only to the thermal circulations of the massive Tharsis volcanoes, but to thermal circulations associated with lesser topography such as crater rims and hills as further emphasized in the statement, "Mesoscale simulations suggest that mass, dust, water and other volatiles can be transported from the boundary layer to heights in excess of 30 km in an hour or less by the thermal circulations associated with the largest topographic features on Mars. Smaller orography produces similar but smaller transport. The net large-scale vertical transport need not (and in fact, may not) be accomplished principally by the Hadley cell circulation."

NLDT on Mars is not limited to dust. The transport of water was specifically targeted in (Michaels et al. 2006). In this paper, it was shown that, like dust, the thermal circulations of the Tharsis volcanoes provided substantial deep, vertical transport of water. Quantitative analysis of model output showed that the vertical flux was roughly one third of the total water flux that would be provided by the Hadley Cell. Thus, the localized upward motion associated with Tharsis and other topographic circulations may very well constitute almost entirely the mean upward motion that results in the mathematically constructed Hadley Cell, just as upward motion in clouds is entirely responsible for the upward rising branch of the terrestrial Hadley

Cell. Or, as stated by Michaels et al. (2006), “Such mountain-induced circulations are thus an important facet of the global water cycle, and possibly the dust cycle as well. This indicates that one longitudinal asymmetry in the MGCM Hadley cell water transport (dominant/enhanced rising branch over the Tharsis region) may in reality be substantially due to these volcano induced circulations.”

Thermal circulations may be the dominant NLDT mechanism on Mars, but the observational evidence for NLDT in dust storms cannot be ignored. Local dust storms occur on Mars on almost a daily basis and larger, regional dust storms are also relatively frequent. Orbital imagery of these disturbances clearly show coherent, convective circulations that rapidly transport dust from the surface into the free troposphere. The veils of dust that linger for days attest to this transport. In simulating idealized dust storms, Rafkin (2009) showed that the convective circulations in some dust storms behave like the thunderstorms within terrestrial tropical storms; the updrafts in dust storms are then analogous to the hot towers of Riehl and Malkus.

Numerous studies on the Mars dust cycle have been published, all of which rely on maintaining the background dust load through a combination of dust devil and mean wind lifting followed by transport via resolved motion and turbulent (and numerical) diffusion (e.g., Newman et al. 2002, Basu et al. 2004, and Kahre et al. 2006). Dust devils are known to lift dust, but these circulations are confined to the planetary boundary layer and are therefore unlikely to contribute directly to dust in the free atmosphere above. Dust devils likely do play an important role in maintaining the boundary layer dust load upon which NLDT mechanisms can feed. Only then to the extent that the models capture NLDT circulations will the modeled dust cycle match reality. This is exactly analogous to the moist and dry Hadley Cell studies described for Earth (Manabe et al. 1970). At present, general circulation models may only partially capture NLDT from the large Tharsis volcanoes and capture none of the NLDT from smaller-scale topographic features or the convective updrafts present within dust storms. Therefore, it is very likely that these models are under representing the flux of dust from the boundary layer into the free atmosphere.

There is nothing special about water or dust. Any quantity that can be advected should respond in a similar manner. This includes other chemical species such as $CH_4$. The concentration and distribution of $CH_4$ is presently an area of strong interest, but no attention has been given to whether chemical species can be rapidly vented and transported from the boundary layer. Likewise, no attention has been given to how NLDT may impact chemical processes as they do in terrestrial clouds. Photochemical models still rely solely on diffusion and Hadley Cell-like motion to transport chemical species (e.g., Shimazaki 1989, Nair et al. 1994, Atreya and Gu 1995).

### 3.2 Titan

The clouds of Titan have been an active area of investigation since confirmation of their existence (Griffith et al. 1998) and concomitant with ongoing observations (e.g., Brown et al., 2002, Roe and et al. 2005, Schaller et al., 2006a, Rannou et al. 2006). The climatology (location and season) of these clouds has been tied to seasonal changes in solar heating and surface heat flux (e.g., Brown et al., 2002, Schaller et al., 2006b) and to the seasonal migrations of the upwelling branch of the presumed global Hadley cell (e.g., Tokano et al. 2001, Mitchell et al. 2006). Barth and Rafkin (2007) explicitly simulated deep convective methane clouds and showed that boundary layer relative humidity values strongly regulated the intensity and depth of the clouds. The general circulation modeling study of Mitchell et al. (2006) also clearly showed that the boundary layer methane humidity, partly controlled by the Hadley cell circulation, is also a key element to understanding cloud climatology.

One element that all these previous studies have in common is that clouds are viewed as independent entities that respond to large-scale thermodynamic and dynamic conditions. However, to the extent that Earth's hydrological cycle serves as an analog to Titan's methane cycle, convective clouds are not just a response to large-scale conditions and circulation. Rather, clouds are part of the circulation, and in the case of the rising branch of Earth's Hadley cell, clouds are the upward circulation.

As on Earth then, convective clouds on Titan may play an important role in climate dynamics, atmospheric chemistry, and the overall volatile cycle. Clouds may regulate the climate of Titan directly, through radiative forcing (enhanced greenhouse or anti-greenhouse), and perhaps through latent heat release. They may also indirectly force the climate by controlling the distribution of radiatively active gases and aerosols; Titan's convective clouds provide a potential mechanism for rapid and deep tropospheric transport in what is thought to be an otherwise sluggish global circulation (*e.g,* Hourdin et al. 1995, Tokano et al. 2001). At the same time, it is the large-scale circulation that creates a conditionally unstable environment, which is a necessary condition for convective clouds. Also, if Titan's clouds are analogous to those on the Earth, the large-scale environment predetermines the nature and character of the convection. Thus, the large-scale circulation and smaller-scale convective clouds are intimately coupled.

That the observed convective clouds of Titan are associated with the rising branch of the Hadley Cell is more than a coincidence. If these clouds behave like the tropical clouds of Earth, the clouds are the Hadley Cell—it could be no other way. And, like Earth, outside the clouds the large-scale atmosphere is likely descending. It is more proper to say that cloud updrafts constitute the rising branch of the Hadley cell and that it is the flux of heat and especially moisture in the poleward surface branch of the Hadley Cell that sustains these clouds.

General circulation model studies of Titan's atmosphere have all neglected deep convection. Tokano et al. (2001) specifically mentioned that the effect of deep convective clouds was neglected, because so little was known about the clouds. Deep convective clouds were ignored altogether in Hourdin et al. (1995) and Rannou et al. (2006), although such clouds were not known to exist in the case of the former study. Mitchell et al. (2006) noted the importance of large-scale methane advection required to support convection, but did not include the convective effects of such clouds.

Although not a GCM study, Mitchell (2009) demonstrated the potential importance of including NLDT by convective clouds. Through a simple analytical model, the atmospheric angular momentum budget of the atmosphere could be substantially modified. The differences between a dry model and a deep convective model were sufficient to produce measurable differences in the spin rate of Titan's surface. Since convective clouds are known to modulate the angular momentum of the Earth's atmosphere, it may be that Titan's convective clouds may also play an important role in superrotation, which has been a long-standing dynamical problem (Del Genio and Zhou 1996, Zhu 2006).

## 4 Summary and Conclusion

NLDT is an important transport mechanism for Earth and it plays a key role in regional and global energy, hydrologic, aerosol and chemical cycles. NLDT is necessary to produce elevated layers (local maxima) of energy, aerosols, and chemical species that are physically separated from their source. Observations of energy, momentum, water, aerosols and other chemical species cannot be reproduced without incorporating NLDT.

Deep convective clouds are the most important and best understood NLDT mechanism, and the primary manifestation is in providing the upward branch of the Hadley Cell. The Hadley Cell is a mathematical construct that plays no physical role in vertical transport within the tropics. Instead, transport is accomplished through the updrafts in the clouds and the compensating subsidence forced by the clouds.

Besides convective clouds, forest fires, dust storms, topographic circulations and volcanic eruptions can also produce NLDT. NLDT processes must be included to properly characterize many global and regional atmospheric cycles on Earth.

There is observational evidence, in the form of elevated dust layers, of NLDT on Mars and NLDT mechanisms (e.g., deep convective clouds) are known to exist on Titan. It is almost inescapable that NLDT is playing an important role in these planetary atmospheres.

On Mars, NLDT of dust is the most likely mechanism by which elevated dust layers are produced. If so, the NLDT hypothesis forwarded by (Rafkin et al. 2002), and (Rafkin 2003) should be included as part of a complete theory of the Mars dust cycle. In particular, while dust devils and mean wind lifting provide a source of dust to the boundary layer, the NLDT mechanism transports this dust out of the boundary layer and into the free atmosphere. The upward branches of these thermal circulations and within dust storms are the physical circulations that actually produce transport, while the upward branch of the Hadley Cell is a mathematical construct just as it is on Earth. NLDT is a reasonable explanation for the observations of elevated layers of dust.

With respect to water and other species, NLDT on Mars will influence transport in a manner similar to dust; water, $CH_4$, $O_3$, etc. can all be vented from the boundary layer via NDLT processes, as described by Michaels et al. (2006). It should, however, be recognized that water and dust cycles can interact, since dust may serve as condensation nuclei. The rapid transport of species such as $CH_4$ could have a dramatic impact on the highly controversial and still poorly understood distribution of $CH_4$, which for now, has been modeled via general circulation models that can only crudely resolve NLDT associated with the largest topographic features (e.g., Lefevre and Forget 2009).

Deep convective clouds on Titan are seasonably sporadic; none of these clouds have been observed as Titan moves through equinox. The clouds may appear in the northern summer high latitudes, but this remains to be seen. NLDT would, therefore, also be seasonally sporadic. As a consequence, vertical transport of quantities like methane would be efficient when deep convective clouds are present, and would be relatively inefficient at other times. If this is the case, a seasonal chemical cycle might be present in the upper troposphere and lower stratosphere.

Under a seasonal chemical cycle scenario, large quantities of methane would be rapidly transported out of the (well-mixed) boundary layer and injected into the upper troposphere and lower stratosphere when deep convective clouds are active in the polar regions. This transport would be substantially greater than might be achieved by a slow Hadley Cell or by diffusion alone. There should be a mid-tropospheric minimum of $CH_4$ outside of the clouds. Once aloft, the methane would be transported equatorward, undergoing photochemical reactions along the way. If the primary source of methane is at the poles, there should be a gradual decrease of methane away from the source region and an increase in methane photolysis byproducts such as ethane. When convection ceases, the NLDT source of methane also ceases. Only by slow Hadley Cell ascent and diffusion could $CH_4$ be supplied to the upper troposphere and lower stratosphere when clouds are not active. Therefore, a seasonal decrease of methane at these levels as well as a decrease in photochemical production rates of byproducts might be observed.

If NLDT is an important process on Mars and Titan, there are tell-tale signs and predictions that would confirm this. Some of this evidence has already been obtained (e.g., elevated dust layers). Other direct evidence for Mars would include observations of enhanced plumes over topographic features, elevated layers and a mid-tropospheric minimum of water in regions of active NLDT, and compensating subsidence in regions of the rising branch of the Hadley Cell. Direct evidence on Titan would be more analogous to that of Earth: tropospheric minimums of moist static energy and $CH_4$, and large-scale compensating subsidence. Numerical models can also play an important role in supporting the NLDT hypothesis. For Earth, inclusion of NLDT processes via parameterizations has led to an improvement of model results when compared with observations. Parameterization of NLDT processes for Mars and Titan might provide similar improvements. Further study on the importance of NLDT in atmospheres other than Earth is warranted.

## Acknowledgements

Mr. Timothy Michaels is acknowledged for his insight during numerous discussions on the topic of deep transport mechanisms. This work was supported by NASAs Planetary Atmosphere Program and the Outer Planets Fundamental Research Program.